\documentclass[twoside,twocolumn,english,prc,amsmath,a4paper,myheadings]{revtex4}
\usepackage[T1]{fontenc}
\usepackage{geometry}
\usepackage{amsmath}
\usepackage{color}
\usepackage{graphicx}
\usepackage{amssymb}

\makeatletter
\def\@oddhead{\rightmark \hfill Evidence for Hydrodynamic Evolution in Proton-Proton Scattering at LHC Energies\hfill \thepage}
\def\@evenhead{\thepage \hfill K. Werner et al.\hfill}
\def\fnum@table{\tablename~{\bf\thetable}}
\def\fnum@figure{\figurename~{\bf\thefigure}}
\def\tablename{\footnotesize{\bf Table}}
\def\figurename{\footnotesize{\bf Figure}}

\usepackage{dcolumn}

\def\citet{\cite}

\makeatother

\usepackage{babel}

\begin{document}

\title{{\normalsize Evidence for Hydrodynamic Evolution in Proton-Proton
Scattering at LHC Energies}}

\author{{\normalsize K.$\,$Werner$^{(a)}$, Iu.$\,$Karpenko$^{(a,b)}$,
T.$\,$Pierog$^{(c)}$, M. Bleicher$^{(d)}$, K. Mikhailov$^{(e)}$}}

\address{$^{(a)}$ SUBATECH, University of Nantes -- IN2P3/CNRS-- EMN, Nantes,
France}

\address{$^{(b)}$ Bogolyubov Institute for Theoretical Physics, Kiev 143,
03680, Ukraine}

\address{$^{(c)}$ Forschungszentrum Karlsruhe, Institut fuer Kernphysik,
Karlsruhe, Germany}

\address{$^{(d)}$ Frankfurt Institute for Advanced Studies (FIAS), \\
Johann Wolfgang Goethe Universitaet, Frankfurt am Main, Germany}

\address{$^{(e)}$ Institute for Theoretical and Experimental Physics, Moscow,
117218, Russia}

\begin{abstract}
In $pp$ scattering at LHC energies, large numbers of elementary scatterings
will contribute significantly, and the corresponding high multiplicity
events will be of particular interest. Elementary scatterings are
parton ladders, identified with color flux-tubes. In high multiplicity
events, many of these flux tubes are produced in the same space region,
creating high energy densities. We argue that there are good reasons
to employ the successful procedure used for heavy ion collisions:
matter is assumed to thermalizes quickly, such that the energy from
the flux-tubes can be taken as initial condition for a hydrodynamic
expansion. This scenario gets spectacular support from very recent
results on Bose-Einstein correlations in $pp$ scattering at 900 GeV
at LHC.
\end{abstract}
\maketitle

\section{Introduction}

After one decade of RHIC experiments it seems to be certain that heavy
ion collisions at RHIC energies produce a new state of matter which
expands as an almost ideal fluid \citet{hydro1d,hydro2,hydro2aa1,hydro2aa2,hydro2b,hydro2e,hydro4b,epos2},
whereas proton-proton scattering is usually considered to be a reference
system, theoretically well under control via perturbative techniques.
Although at very high energy, hadrons experience multiple scatterings
when they hit protons or neutrons, inclusive cross sections calculations
becomes quite simple due to the fact that different multiple scattering
contributions cancel due to destructive interference (AGK cancellations).
The corresponding formulas are simple and can be expressed in terms
of parton distributions functions, based on evolutions equations.

However, in particular at LHC energies where we expect large numbers
of scatterings to contribute significantly, it becomes interesting
to study event classes corresponding to a large number of scatterings
(in practice: high multiplicity events). Here, one needs partial cross
sections, corresponding to a particular multiple scattering type (single,
or double, or triple...). Gribov-Regge theory provides a solution,
in particular when energy sharing is properly taken into account,
as in the EPOS approach.

High multiplicity events are very interesting for the following reasons:
in EPOS for example, a single scattering amounts to the exchange of
a complete parton ladder, including initial state radiation. The whole
object is identified as a pair of color flux tubes, which finally
break into many pieces (hadrons). In high multiplicity events, with
many scatterings involved, we have many partons ladders participating,
and therefore a large number of flux tubes sitting essentially on
top of each other -- as in heavy ion scattering at RHIC. In the heavy
ion case, we simply compute the energy density corresponding to these
flux tubes (from string theory), assume thermalization, and then perform
a hydrodynamic expansion based on these initial conditions \citet{epos2}.

Since the energy densities reached in high multiplicity proton-proton
collisions are comparable to the ones achieved in gold-gold scattering
at RHIC, we will apply the same procedure. The usual argument against
this approach is the small size of the $pp$ system, but since we
know by now that the size of the space fluctuations in an even-by-event
treatment in AuAu scattering is of the order of 1-2 fm, and AuAu seems
to be driven by hydrodynamic flow, there is no reason not to do so
for high multiplicity $pp$.

In this paper, we will briefly review the flux-tube/hydro approach
of \citet{epos2}, with special emphasis on $pp$ scattering. After
some elementary checks concerning particle distributions, we come
to the main result of this paper: the hydrodynamic expansion modifies
drastically the space-time behavior of the evolution, compared to
basic picture where the flux-tubes decay independently. And this space-time
structure can be clearly {}``seen'' when investigating Bose-Einstein
correlations, and the recently published results from ALICE confirm
the {}``hydrodynamic scenario''.

\section{Multiple Scattering}

\subsection{Parton evolution}

\begin{figure}[tb]
\begin{centering}
\includegraphics[scale=0.45]{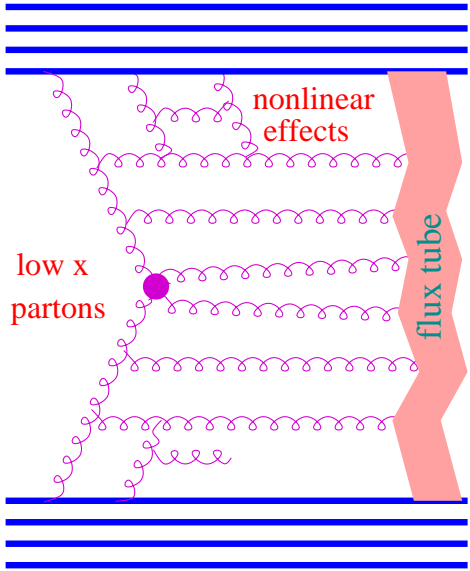}
\par\end{centering}

\caption{(Color online) Elementary interaction in the EPOS model.\label{cap:Elementary-interaction}}

\end{figure}

An elementary scattering within the EPOS approach \citet{epos2} is
given by a so-called {}``parton ladder'', see fig. \ref{cap:Elementary-interaction},
\textbf{\large }%
representing parton evolutions from the projectile and the target
side towards the center (small $x$). The evolution is governed by
an evolution equation, in the simplest case according to DGLAP. In
the following we will refer to these partons as {}``ladder partons'',
to be distinguished from {}``spectator partons''. Such a parton
ladder may be considered as a longitudinal color field or flux-tube,
conveniently treated as a relativistic string. The intermediate gluons
are treated as kink singularities in the language of relativistic
strings. This flux tube approach is just a continuation of 30 years
of very successful applications of the string picture to particle
production in collisions of high energy particles \citet{and83,wer93,cap94,nexus},
in particular in connection with the parton model. An important issue
at high energies is the appearance of so-called non-linear effects,
which means that the simple linear parton evolution is no longer valid,
gluon ladders may fuse or split. More recently, a classical treatment
has been proposed, called Color Glass Condensate (CGC), having the
advantage that the framework can be derived from first principles
\citet{cgc1,cgc2,cgc3,cgc4,cgc5}. Comparing a conventional string
model like EPOS and the CGC picture: they describe the same physics,
although the technical implementation is of course different. All
realistic string model implementations have nowadays to deal with
screening and saturation, and EPOS is not an exception, see \citet{kw-split,epos2}.
Without screening, proton-proton cross sections and multiplicities
will explode at high energies. 

\begin{figure}[tb]
\begin{centering}
{\large (a)}\hspace*{7cm}~
\par\end{centering}

\begin{centering}
\vspace*{-0.5cm}\includegraphics[scale=0.45]{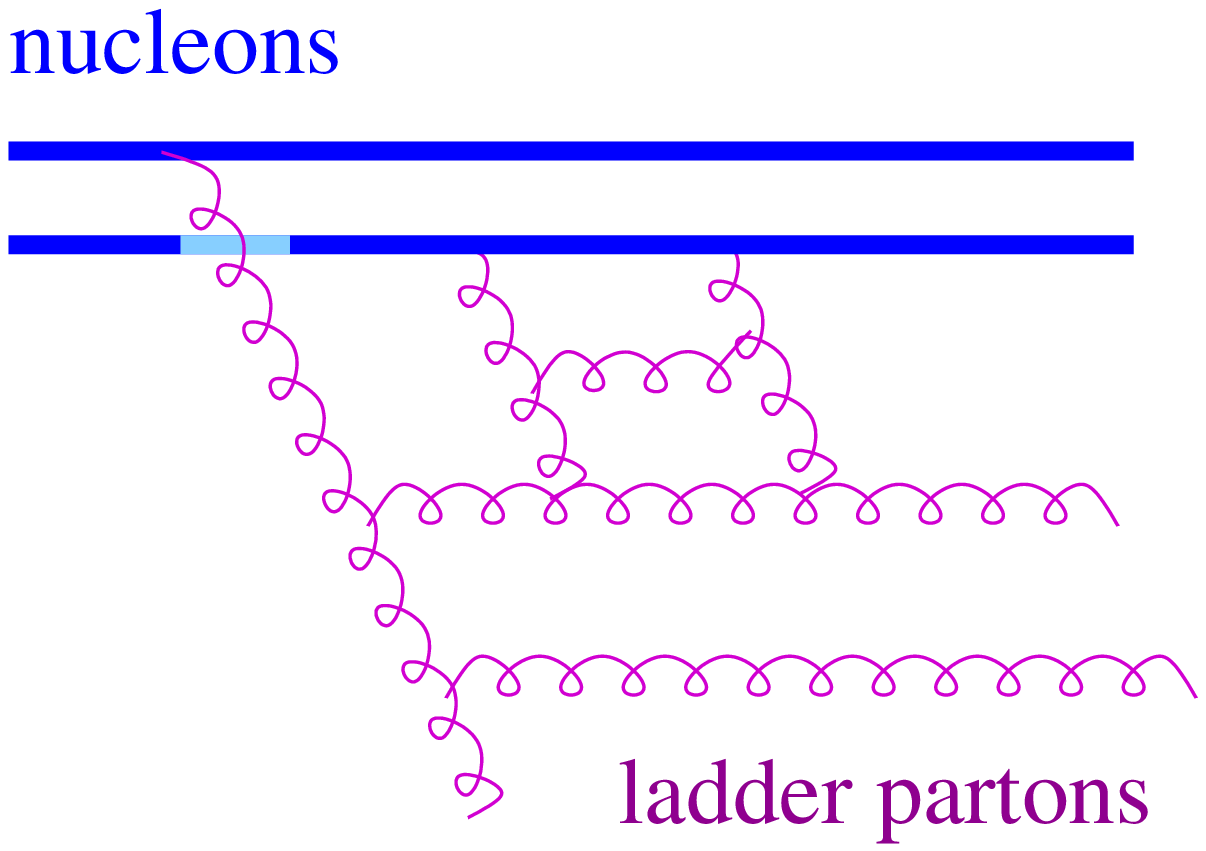}
\par\end{centering}

\begin{centering}
{\large (b)}\hspace*{7cm}~
\par\end{centering}

\begin{centering}
\vspace*{-0.5cm}\includegraphics[scale=0.45]{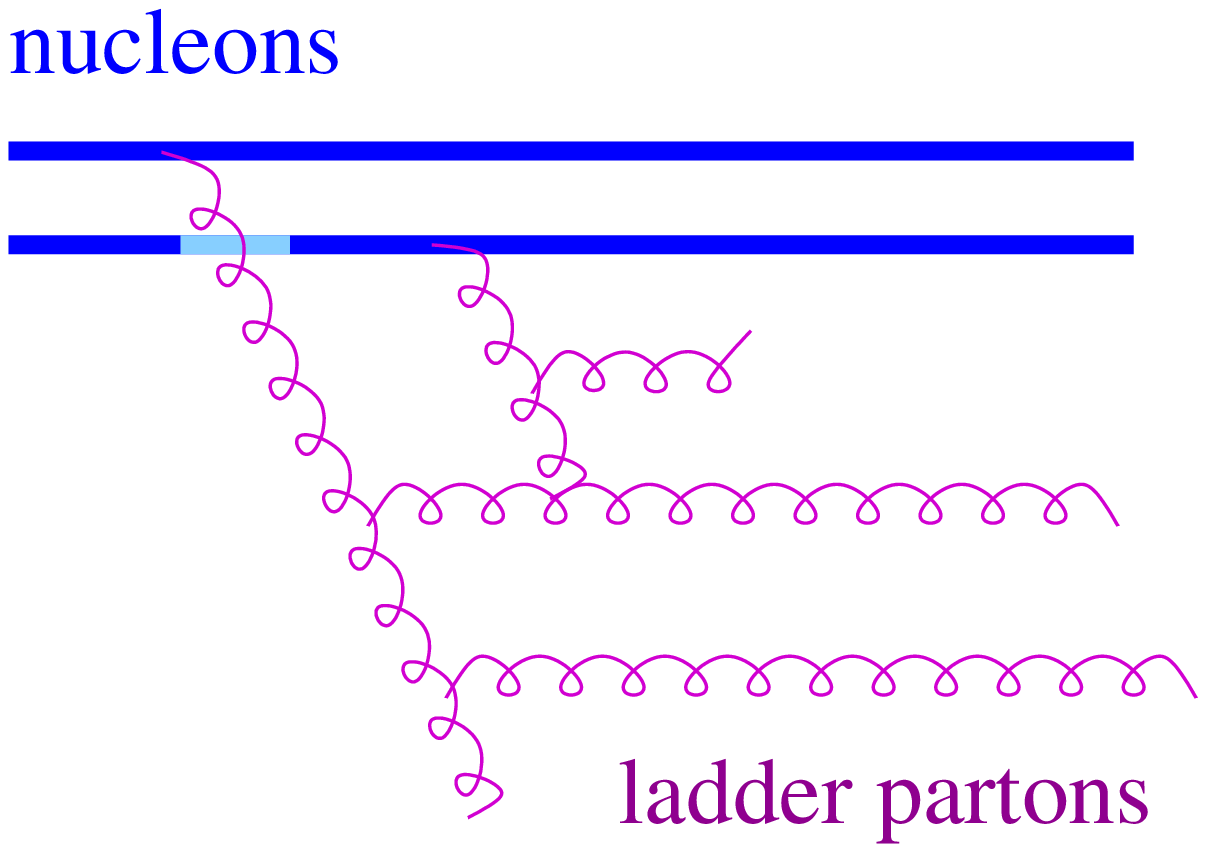}
\par\end{centering}

\caption{(Color online) (a) Elastic {}``rescattering'' of a ladder parton.
We refer to elastic parton ladder splitting; (b) Inelastic {}``rescattering''
of a ladder parton. We refer to inelastic parton ladder splitting.\label{split}}

\end{figure}

A phenomenological treatment of non-linear effects in EPOS employs
two contributions: a simple elastic rescattering of a ladder parton
on a projectile or target nucleon (elastic ladder splitting), or an
inelastic rescattering (inelastic ladder splitting), see fig. \ref{split}.
The elastic process provides screening, therefore a reduction of total
and inelastic cross sections. The importance of this effect should
first increase with mass number (in case of nuclei being involved),
but finally saturate. The inelastic process will affect particle production.
Both, elastic and inelastic rescattering must be taken into account
in order to obtain a realistic picture.%

To include the effects of elastic rescattering, we first parametrize
a parton ladder (to be more precise: the imaginary part of the corresponding
amplitude in impact parameter space) computed on the basis of DGLAP.
We obtain an excellent fit of the form $\alpha(x^{+}x^{-})^{\beta}$,
where $x^{+}$ and $x^{-}$ are the momentum fractions of the {}``first''
ladder partons on respectively projectile and target side (which initiate
the parton evolutions). The parameters $\alpha$ and $\beta$ depend
on the cms energy $\sqrt{s}$ of the hadron-hadron collision. To mimic
the reduction of the increase of the expressions $\alpha(x^{+}x^{-})^{\beta}$
with energy, we simply replace them by \begin{equation}
\alpha(x^{+})^{\beta+\varepsilon_{P}}(x^{-})^{\beta+\varepsilon_{T}},\end{equation}
where the values of the positive numbers $\varepsilon_{P/T}$ will
increase with the nuclear mass number and $\log s$.

The inelastic rescatterings (ladder splittings, looking from inside
to outside) amount to providing several ladders close to the projectile
(or target) side, which are close to each other in space. They cannot
be considered as independent color fields (strings), we should rather
think of a common color field built from several partons ladders.
We treat this object via an enhancement of remnant excitations. In
fact, the picture described so far is not yet complete, since we just
considered two interacting partons, one from the projectile and one
from the target. Also the remnants themselves contribute to particle
production, but mainly in the fragmentation region. For more details
see \citet{epos2}.

\subsection{Factorization and Multiple Scattering}

An inclusive cross section is one of the simplest quantities to characterize
particle production. As discussed earlier, inclusive cross section
are particularly simple, quantum interference helps to provide simple
formulas referred to a {}``factorization''. If we want to study
high multiplicity events, we have to go beyond the inclusive treatment.

To formulate a consistent multiple scattering theory is difficult.
A possible solution is Gribov's Pomeron calculus, which can be adapted
to our language by identifying Pomeron and parton ladder. Multiple
scattering means that one has contributions with several parton ladders
in parallel. This formulation is equivalent to using the eikonal formula
to obtain the total cross section from the knowledge of the inclusive
one.

\textbf{\large }%
\begin{figure}[b]
\begin{centering}
\includegraphics[scale=0.4]{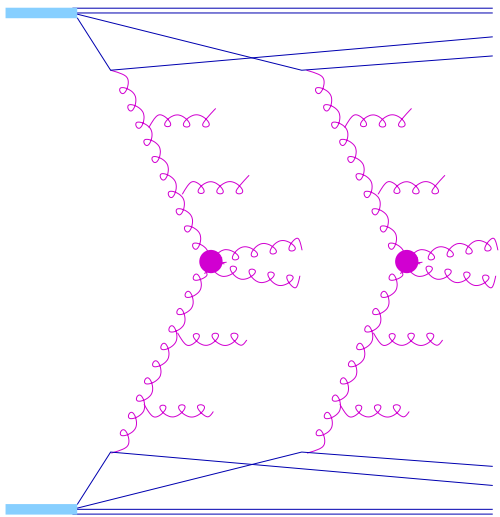}
\par\end{centering}

\caption{(Color online) Multiple scattering with energy sharing.\label{cap:Energy}}

\end{figure}
We indicated several years ago inconsistencies in this approach, proposing
an {}``energy conserving multiple scattering treatment'' \citet{nexus}.
The main idea is simple: in case of multiple scattering, when it comes
to calculating partial cross sections for double, triple ... scattering,
one has to explicitly care about the fact that the total energy has
to be shared among the individual elementary interactions. In other
words, the partons ladders which happen to be parallel to each other
share the collision energy, see fig. \ref{cap:Energy}. A consistent
quantum mechanical formulation of these ideas requires not only the
consideration of the usual (open) parton ladders, discussed so far,
but also of closed ladders, representing elastic scattering. These
are the same closed ladders which we introduced earlier in connection
with elastic rescatterings. The closed ladders do not contribute to
particle production, but they are crucial since they affect substantially
the calculations of partial cross sections. Actually, the closed ladders
simply lead to large numbers of interfering contributions for the
same final state, all of which have to be summed up to obtain the
corresponding partial cross sections. It is a unique feature of our
approach to consider explicitly energy-momentum sharing at this level
(the {}``E'' in the name EPOS). For more details see \citet{nexus}.

\section{Hydrodynamic Evolution}

\subsection{Parton-Ladders, Flux-Tubes, Energy-Momentum Tensor}

\begin{figure}[tb]
\vspace*{0.2cm}

\begin{centering}
{\large (a)}\hspace*{7cm}~
\par\end{centering}

\begin{centering}
\vspace*{0.2cm}\includegraphics[scale=0.33]{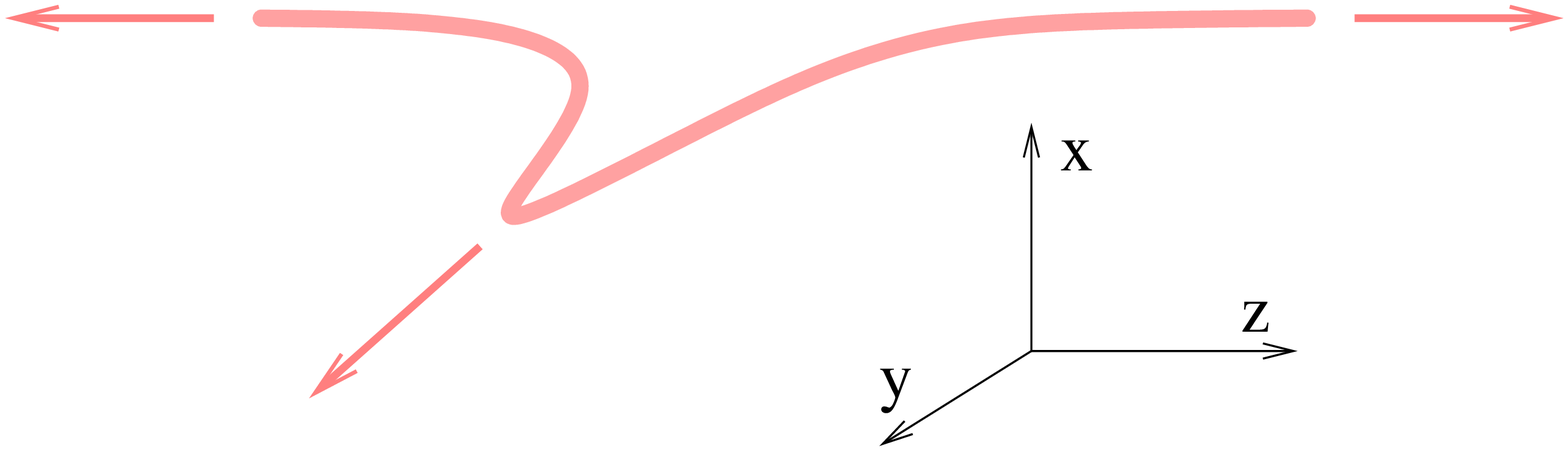}
\par\end{centering}

\begin{centering}
{\large (b)}\hspace*{7cm}~
\par\end{centering}

\begin{centering}
\vspace*{0.2cm}\includegraphics[scale=0.33]{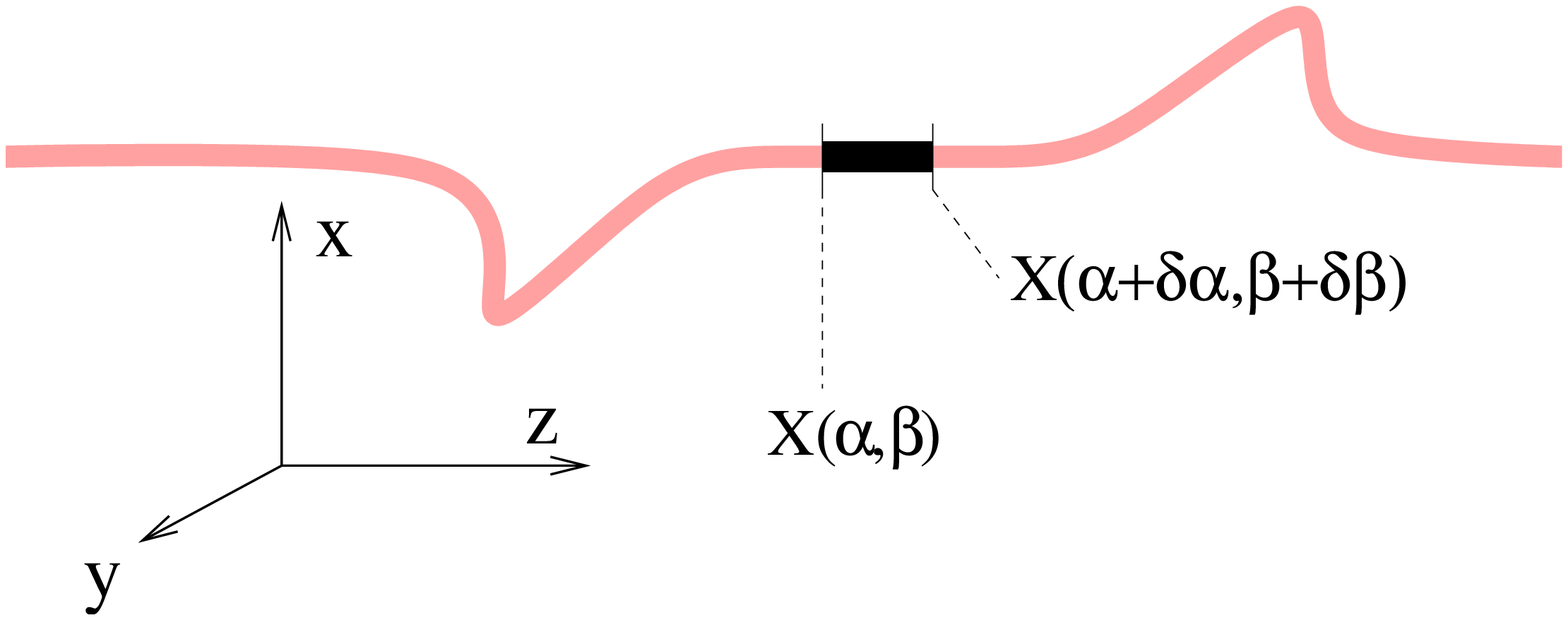}
\par\end{centering}

\caption{(Color online) (a) Flux tube with transverse kink in $\mathrm{I\! R^{3}}$
space. The kink leads to transversely moving string regions (transverse
arrow). (b) String segment at given proper time. \label{cap:geom0b}}

\end{figure}

In case of high multiplicity $pp$ scattering, we apply exactly the
same procedure as we did for AuAu collisions at RHIC, as explained
in detail in \citet{epos2}, and shortly reviewed in the following.
We will identify parton ladders with elementary flux tubes, the latter
ones treated as classical strings. We use the simplest possible string:
a two-dimensional surfaces $X(\alpha,\beta)$ in 3+1 dimensional space-time,
with piecewise constant initial conditions, referred to as kinky strings.
In fig. \ref{cap:geom0b}(a), we sketch the space components of this
object: the string in $\mathrm{I\! R^{3}}$ space is a mainly longitudinal
object (here parallel to the $z$-axis) but due to the kinks (associated
to transversely moving gluons) there are string pieces moving transversely
(in $y$-direction in the picture). But despite these kinks, most
of the string carries only little transverse momentum!

In case of elementary reactions like electron-positron annihilation
or proton proton scattering (at moderately relativistic energies),
hadron production is realized via string breaking, such that string
fragments are identified with hadrons. When it comes to heavy ion
collisions or very high energy proton-proton scattering, the procedure
has to be modified, since the density of strings will be so high that
they cannot possibly decay independently. For technical reasons, we
split each string into a sequence of string segments, at a given proper-time
$\tau_{0}$, corresponding to widths $\delta\alpha$ and $\delta\beta$
in the string parameter space (see fig. \ref{cap:geom0b}(b)). One
distinguishes between string segments in dense areas (more than some
critical density $\rho_{0}$ of segments per unit volume), from those
in low density areas. The high density areas are referred to as core,
the low density areas as corona \citet{kw-core}. String segments
with large transverse momentum (close to a kink) are excluded from
the core. Based on the four-momenta of infinitesimal string segments,
\textcolor{black}{\begin{equation}
\delta p=\left\{ \frac{\partial X(\alpha,\beta)}{\partial\beta}\delta\alpha+\frac{\partial X(\alpha,\beta)}{\partial\alpha}\delta\beta\right\} ,\end{equation}
}with $g$ being a Gaussian smoothing kernel, one computes the energy-momentum
tensor and conserved currents. The corresponding energy density $\varepsilon(\tau_{0},\vec{x})$
and the flow velocity $\vec{v}(\tau_{0},\vec{x})$ serve as initial
conditions for the subsequent hydrodynamic evolutions. %
\begin{figure}[b]
\begin{centering}
\includegraphics[scale=0.38]{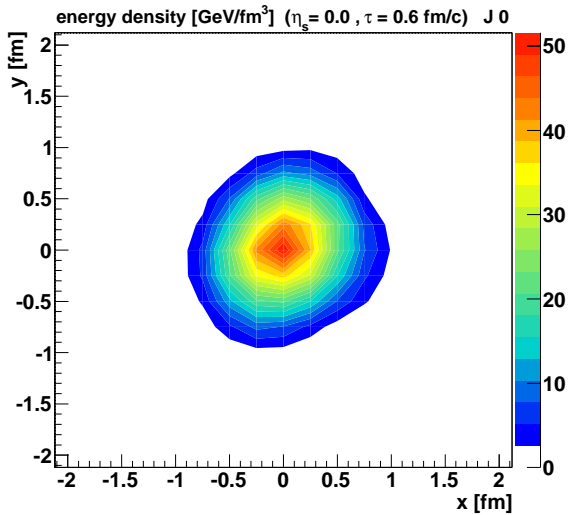}
\par\end{centering}

\caption{(Color online) Initial energy density in a high multiplicity $pp$
collision ($dn/d\eta=12.9$) at 900 GeV, at a space-time rapidity
$\eta_{s}=0$. \label{cap:eiau1}}

\end{figure}

In fig. \ref{cap:eiau1}, we show as an example the energy density
at $\tau_{0}=0.6\,$fm/c for a high multiplicity $pp$ collision at
900 GeV, where high multiplicity here refers to a plateau height $dn/d\eta$
of 12.9, which is more than 3 times the average. We see a maximum
energy density of about 50 GeV/fm$^{3}$, which indeed correspond
to the energy densities observed in central gold-gold collisions at
200 GeV. Even more, comparing with the spiky single event results
for gold-gold in \citet{epos2}, our $pp$ distribution correspond
to one (of many) spikes in gold-gold at 200 GeV, which means a hydrodynamic
treatment for $pp$ is as good (or bad) as for gold-gold at 200GeV.

\subsection{Collective expansion}

Having fixed the initial conditions, matter evolves according to the
equations of ideal hydrodynamics, namely the local energy-momentum
conservation \begin{equation}
\partial_{\mu}T^{\mu\nu}=0,\quad T^{\mu\nu}=(\epsilon+p)\, u^{\mu}u^{\nu}-p\, g^{\mu\nu}\,,\end{equation}
and the conservation of net charges,\begin{equation}
\partial N_{k}^{\mu}=0,\quad N_{k}^{\mu}=n_{k}u^{\mu},\,,\end{equation}
with $k=B,S,Q$, where $B$, $S$, and $Q$ refer to respectively
baryon number, strangeness, and electric charge, and with $u$ being
the four-velocity of the local rest frame. Solving the equations,
as discussed in the appendix of \citet{epos2}, provides the evolution
of the space-time dependence of the macroscopic quantities energy
density $\varepsilon(x)$, collective flow velocity $\vec{v}(x)$,
and the net flavor densities $n_{k}(x)$. %
\begin{figure}[b]
\begin{centering}
\includegraphics[angle=270,scale=0.32]{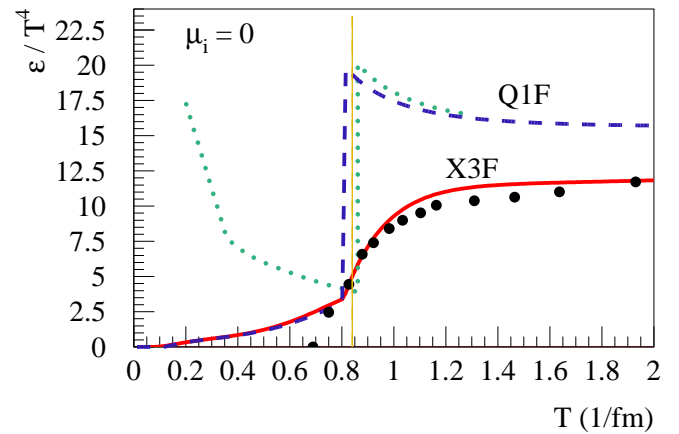}
\par\end{centering}

\caption{(Color online) Energy density versus temperature, for our equation-of-state
X3F (full line), compared to lattice data \citet{lattice} (points),
and some other EoS choices, see \citet{epos2}.\label{cap:eos1} The
thin vertical line indicates the {}``hadronization temperature''
$T_{H}$, i.e. end of the thermal phase, when {}``matter'' is transformed
into hadrons.}

\end{figure}
\begin{figure*}[tb]
\begin{centering}
\includegraphics[scale=0.38]{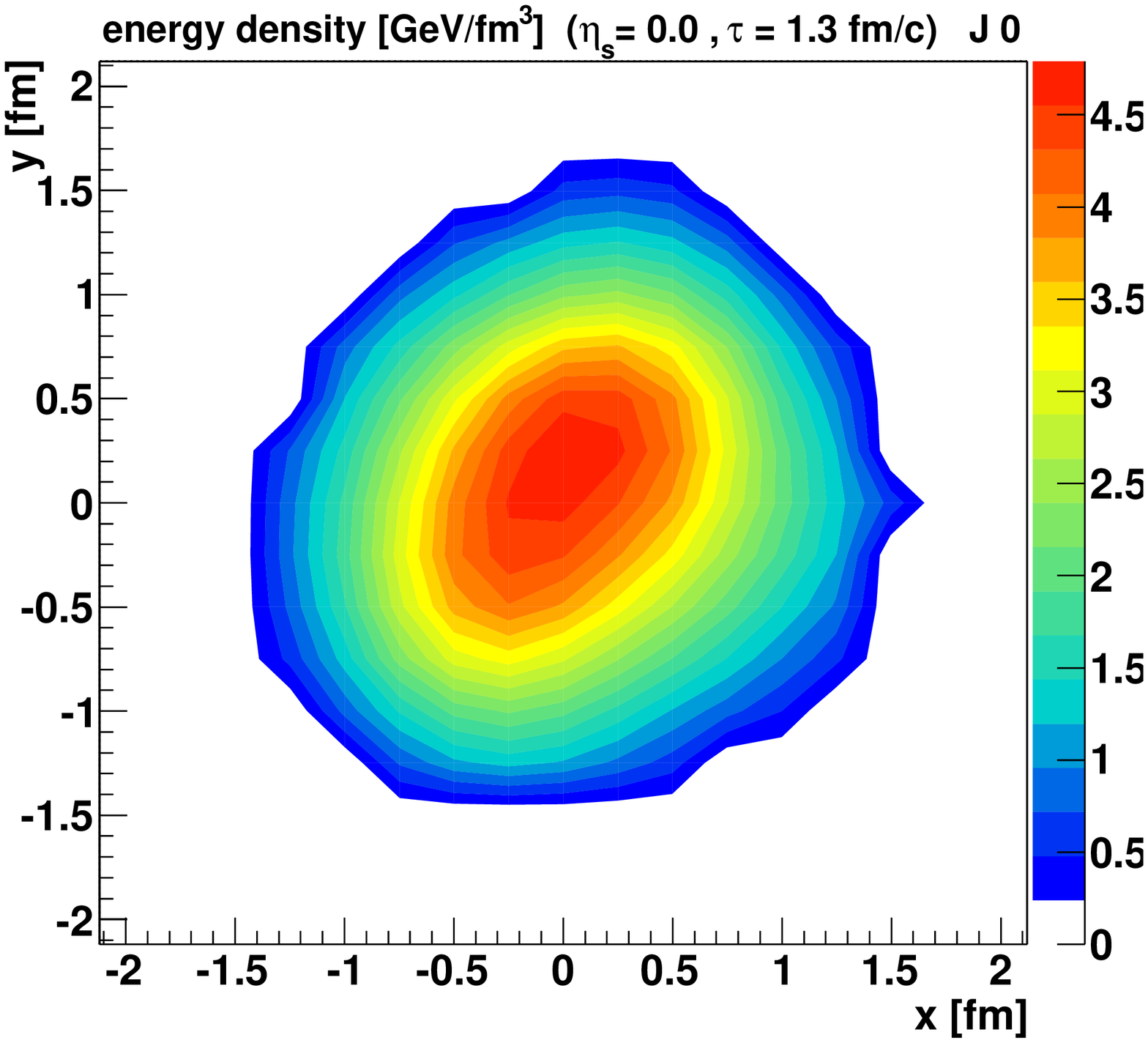}\includegraphics[scale=0.38]{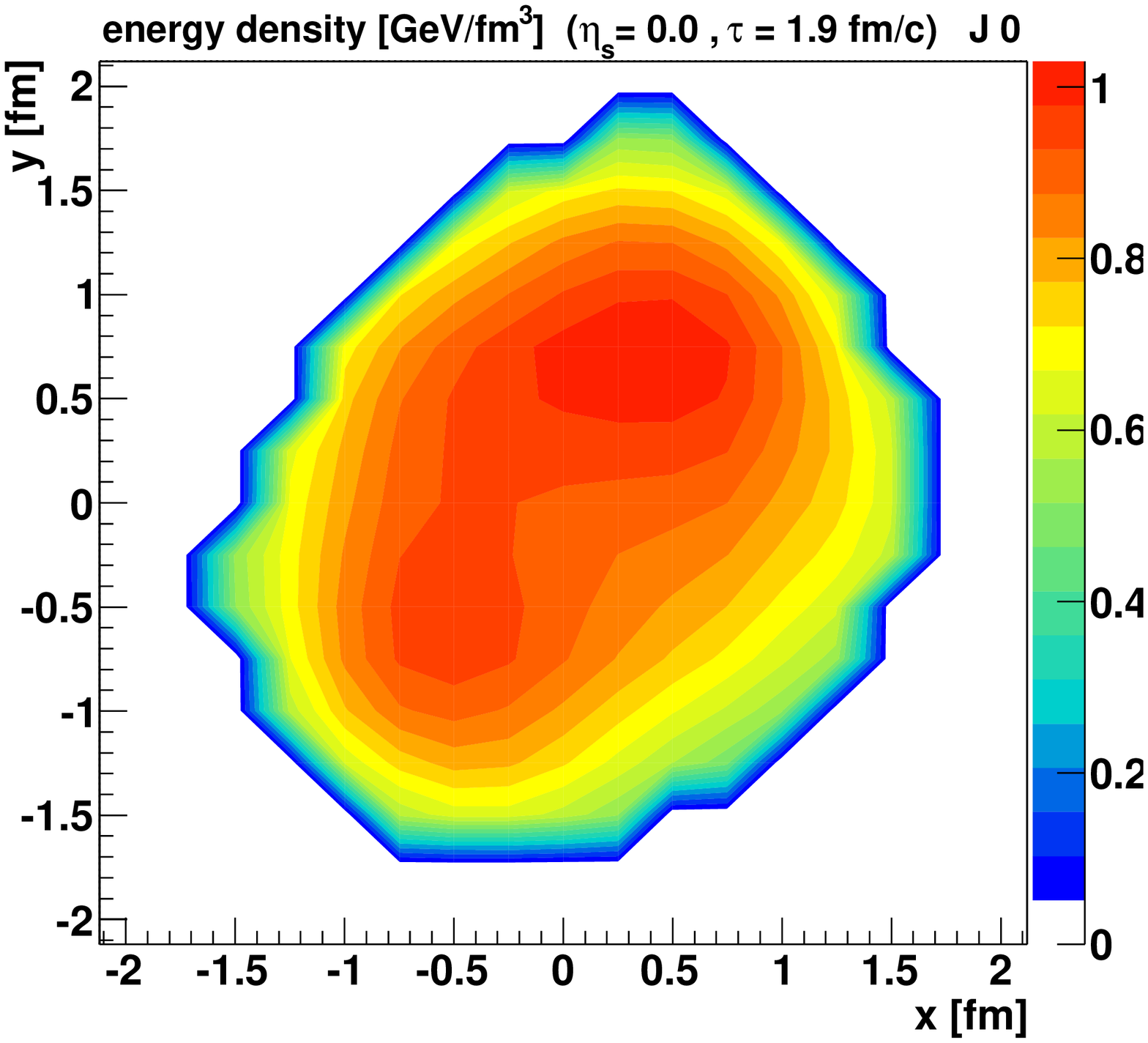}\\
\includegraphics[scale=0.38]{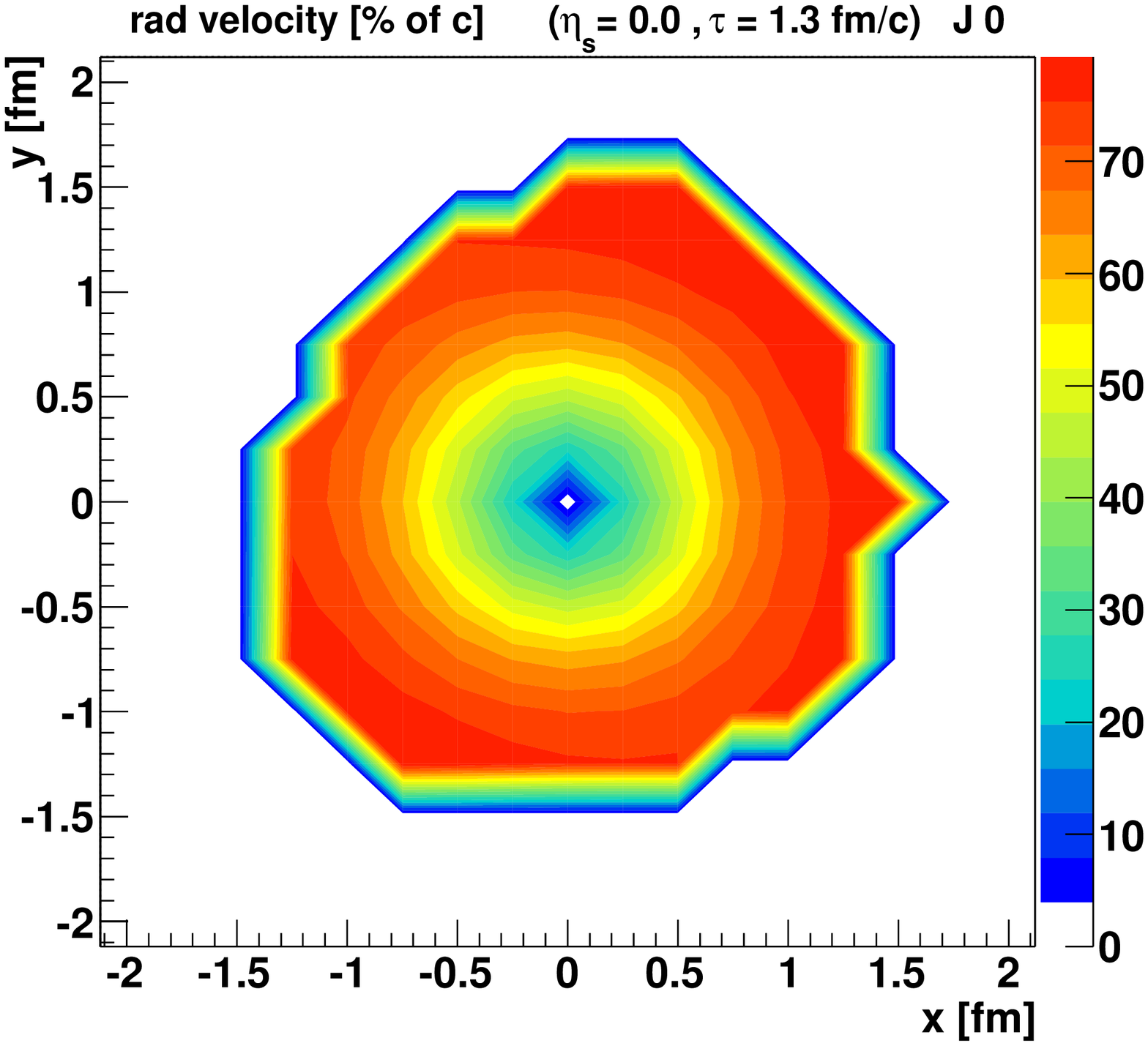}\includegraphics[scale=0.38]{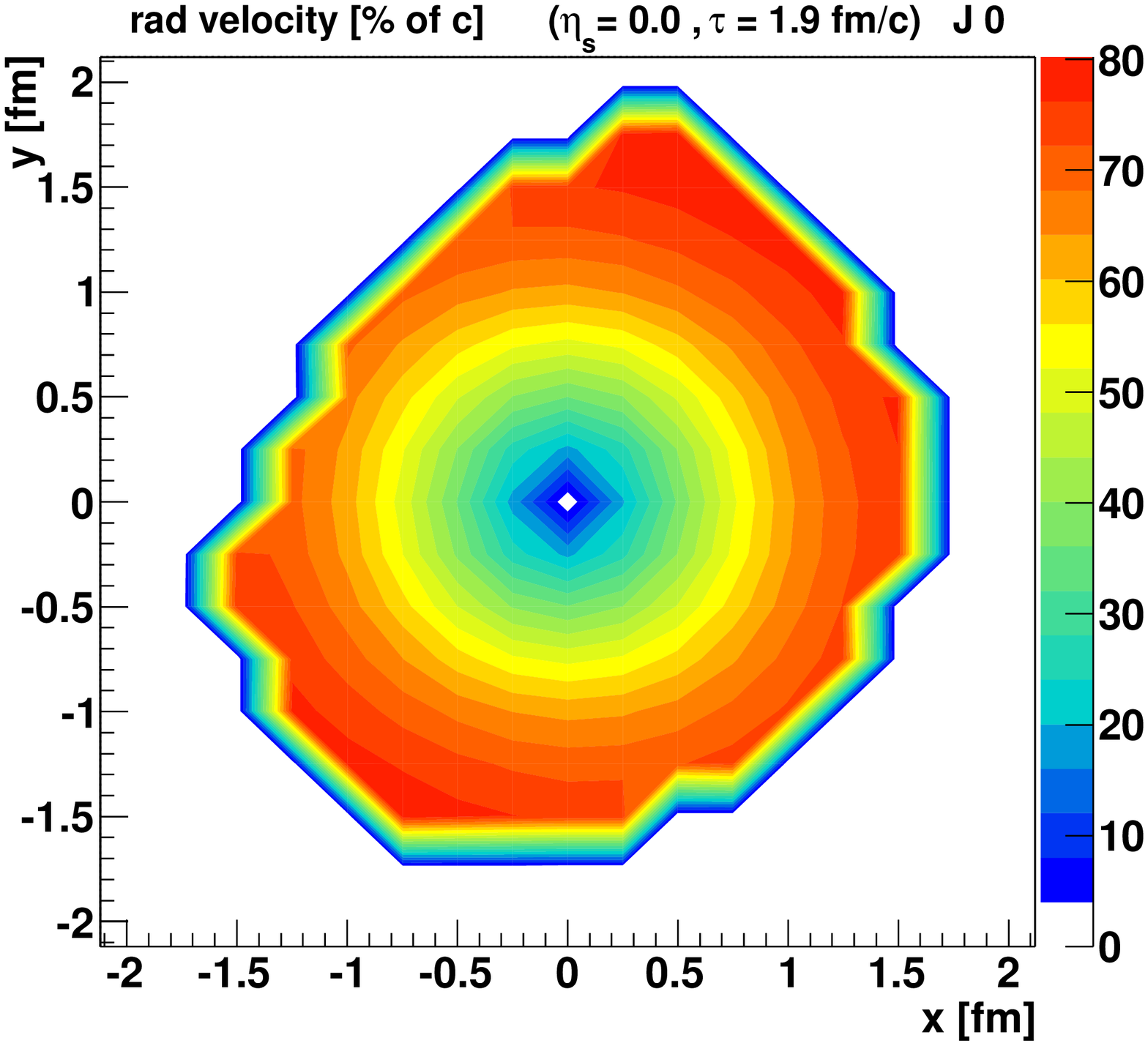}
\par\end{centering}

\caption{(Color online) Energy density (upper panel) and radial flow velocity
(lower panel) for a high multiplicity $pp$ collision ($dn/d\eta=12.9$)
at 900 GeV, at proper times $\tau=1.3\,$fm/c (left) and $\tau=1.9\,$fm/c
(right), at a space-time rapidity $\eta_{s}=0$.\label{cap:eiau3}}

\end{figure*}
Here, the crucial ingredient is the equation of state, which closes
the set of equations by providing the $\varepsilon$-dependence of
the pressure $p$. As discussed in \citet{epos2}, we use an equation
of state compatible with lattice gauge simulations, see fig.~\ref{cap:eos1}.

Starting from the flux-tube initial condition, the system expands
very rapidly. It hadronizes in the cross-over region, where here {}``hadronization''
is meant to be the end of the completely thermal phase: matter is
transformed into hadrons. We stop the hydrodynamical evolution at
this point, but particles are not yet free. Our favorite hadronization
temperature is 166 MeV, shown as the thin vertical line in fig. \ref{cap:eos1},
which is indeed right in the transition region, where the energy density
varies strongly with temperature. At this point we employ statistical
hadronization, which should be understood as hadronization of the
quark-gluon plasma state into a hadronic system, at an early stage,
not the decay of a resonance gas in equilibrium.

After this hadronization --although no longer thermal-- the system
still interacts via hadronic scatterings. The particles at their hadronization
positions (on the corresponding hypersurface) are fed into the hadronic
cascade model UrQMD \citet{urqmd,urqmd2}, performing hadronic interactions
until the system is so dilute that no interactions occur any more.
The {}``final'' freeze out position of the particles is the last
interaction point of the cascade process, or the hydro hadronization
position, if no hadronic interactions occurs.

In fig. \ref{cap:eiau3}, we show the hydrodynamic evolution of the
event corresponding to the initial energy density of fig. \ref{cap:eiau1},
which can be considered as a typical example, with similar observations
being true for randomly chosen events of this multiplicity ($dn/d\eta=12.9$).
We see that the system evolves immediately also transversely, the
energy density drops very quickly. A very large transverse flow develops,
typically around 70 \% of the velocity of light. This will have measurable
consequences.

\section{Elementary distributions}

We first check some elementary distributions. We use the EPOS 2.05
version, which has been optimized for heavy ion scattering at RHIC,
the same one as used in \citet{epos2}. We could certainly improve
the results by doing some {}``tuning'' taking into account the new
LHC results, but the purpose of this paper is more to show what we
get from a straight%
\begin{figure}[b]
\begin{centering}
\includegraphics[angle=270,scale=0.32]{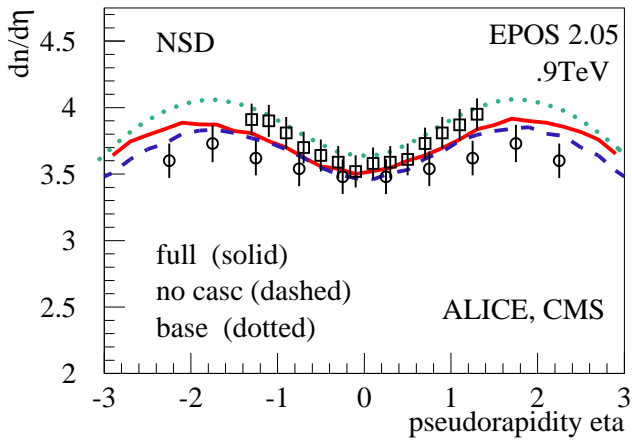}\\

\par\end{centering}

\caption{(Color online) Pseudorapidity distributions in $pp$ scattering at
900 GeV, compared to data (points).\label{cap:rap}We show the full
calculation (solid line), a calculation without hadronic cascade (dashed),
and a calculation without hydro and without cascade (dotted).}

\end{figure}
\begin{figure}[t]
\begin{centering}
\includegraphics[angle=270,scale=0.32]{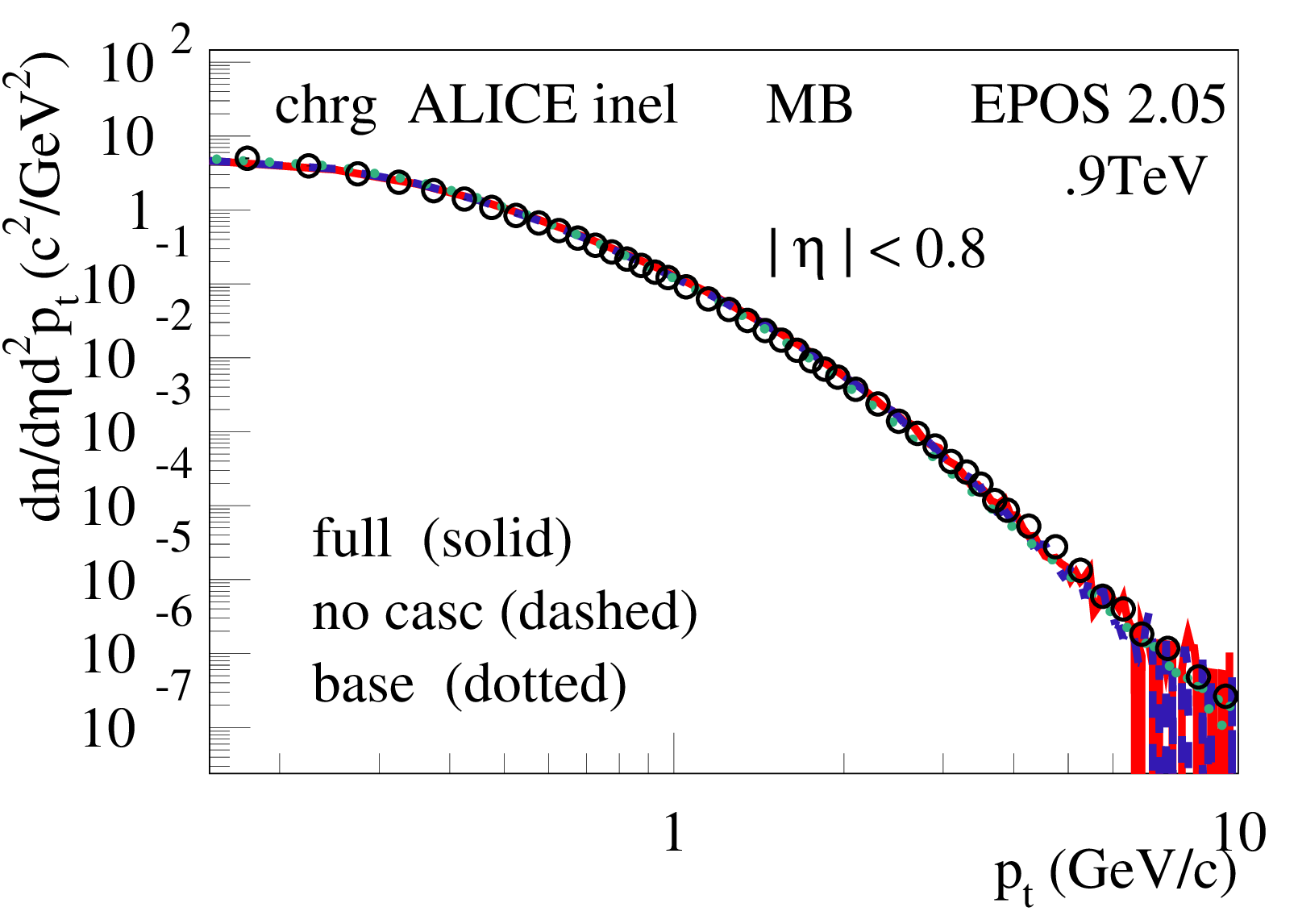}\\
\includegraphics[angle=270,scale=0.32]{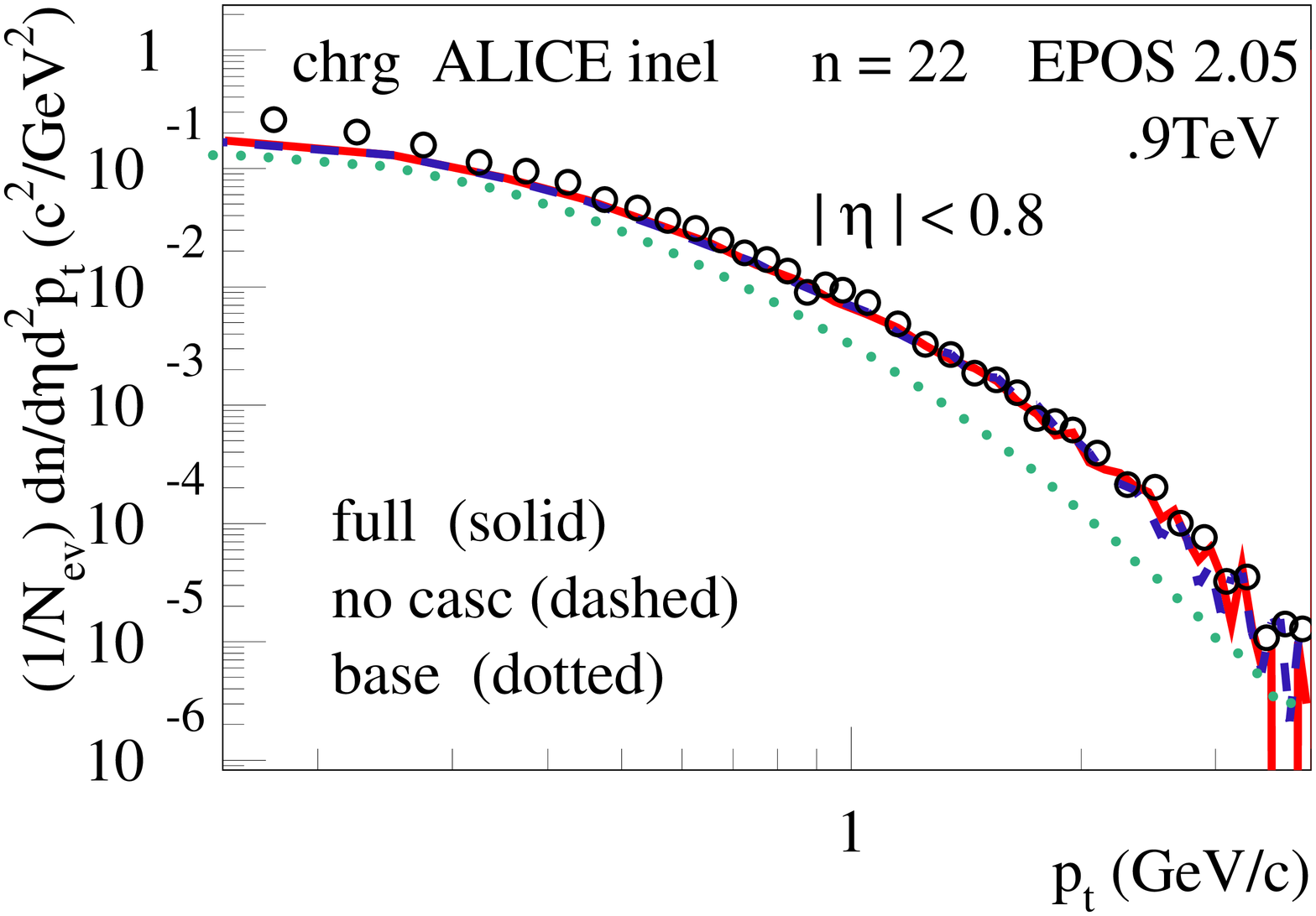}
\par\end{centering}

\caption{(Color online) Transverse momentum distributions in $pp$ scattering
at 900 GeV, for minimum bias events (upper panel) and high muliplicity
events ($n=22$, lower panel), compared to data (points).\label{cap:pt}We
show the full calculations (solid lines), a calculation without hadronic
cascade (dashed), and a calculation without hydro and without cascade
(dotted).}

\end{figure}
\begin{figure}[tb]
\begin{centering}
\includegraphics[angle=270,scale=0.32]{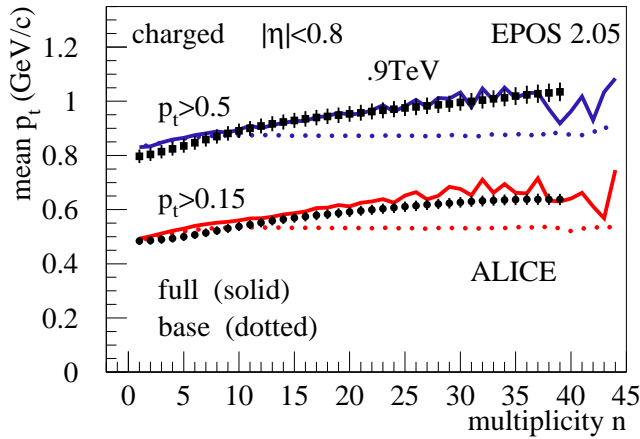}
\par\end{centering}

\caption{(Color online) Mean transverse momentum as a function of the charged
multiplicity in $pp$ scattering at 900 GeV, compared to data (points).
\label{cap:meanpt} We show the full calculation (solid line), and
a calculation without hydro and without cascade (dotted).}

\end{figure}
application of the {}``heavy ion model'', here applied to $pp$
at LHC. We only consider 900 GeV, for higher energies some reconsideration
of our screening procedures will be necessary (work in progress).
As usual we work with the event-by-event mode, and hydrodynamics is
only employed for high density areas (core-corona separation).

In the following we will compare three different scenarios:

\begin{description}
\item [{\underbar{full}}] : the full calculations, including hydro evolution
and hadronic cascade;
\item [{\underbar{no~casc}}] : calculation without hadronic cascade;
\item [{\underbar{base}}] ; calculation without hydro and without cascade.
\end{description}
We will compare the corresponding calculations with experimental data,
for $pp$ scattering at 900 GeV.

In fig. \ref{cap:rap}, we show pseudorapidity distributions of charged
particles, compared to data from CMS \citet{cms} and ALICE \citet{alice1,alice2}.
The three scenarios do not differ very much, and agree roughly with
the data.

We then investigate transverse momentum distributions. For minimum
bias events, there is again little difference for the three scenarios
(all of them reproduce the data within 20\%), as seen in the upper
panel of fig. \ref{cap:pt}. The situation changes drastically, when
we consider high multiplicity events, see the lower panel of fig.
\ref{cap:pt}. Here the base calculation (without hydro) underestimates
the data by a factor of three, whereas the full calculation gets close
to the data. This is a very typical behavior of collective flow: the
distributions get harder at intermediate values of $p_{t}$ (around
1-4 GeV/$c$). 

In fig. \ref{cap:meanpt}, we plot the mean transverse momentum as
a function of the charged multiplicity, compared to data from ALICE
\citet{alice2}. The increase of the mean $p_{t}$ with multiplicity
is in our approach related to collective flow: with increasing multiplicity
one gets higher initial energy densities, and more collective flow
can develop. The data are therefore compatible with our flow picture,
but for a real proof one needs at least in addition the mean $p_{t}$behavior
of heavier particles (protons, lambdas, or even heavier), since the
effect gets bigger with increasing mass.

\section{Bose-Einstein Correlations}

\begin{figure}[b]
\begin{centering}
\hspace*{-0.4cm}\includegraphics[angle=270,scale=0.32]{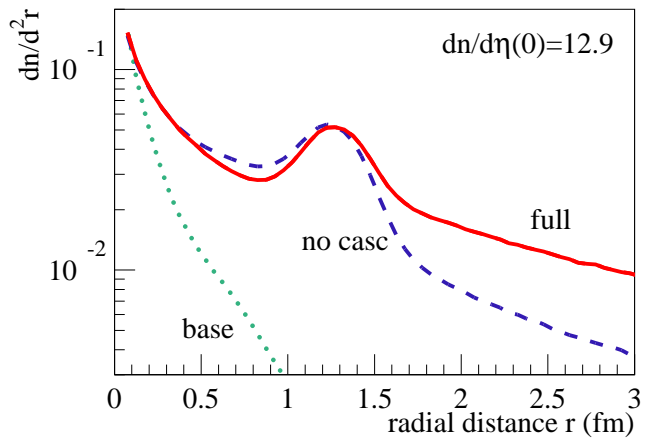}
\par\end{centering}

\begin{centering}
\caption{(Color online) The distribution of formation points of $\pi^{+}$
as a function of the radial distance in a high multiplicity event
from $pp$ scattering at 900 GeV, for the following scenarios: the
full calculation (solid line), a calculation without hadronic cascade
(dashed), and a calculation without hydro and without cascade (dotted).
\label{cap:formation}}

\par\end{centering}
\end{figure}

The space-time evolution of the {}``full'' hydrodynamic approach
will be completely different compared to the {}``base'' approach,
where particles are directly produced from breaking strings, as can
be seen from fig. \ref{cap:formation}, %
where we plot the distribution of formation points of $\pi^{+}$ as
a function of the radial distance \begin{equation}
r=\sqrt{x^{2}+y^{2}}\end{equation}
 (in the $pp$ center of mass system (cms)). Only particles with space-time
rapidities around zero are considered. We compare aain the three scenarios
{}``full'' (full calculation - flux-tube initial conditions, hydro,
hadronic cascade), {}``no~casc'' (without hadronic cascade, only
flux-tube initial conditions and hydro, hadronization as usual at
166 MeV), and {}``base'' (without hydro and without cascade, just
flux-tube approach with string decay). 

All calculations in this section refer to high multiplicity events
in $pp$ scattering at 900 GeV, with a mean $dn/d\eta(0)$ equal to
12.9. The {}``base calculation'' (dotted line) gives as expected
a steeply falling distribution as a function of $r$. In the two cases
involving a hydrodynamical evolution, particle production is significantly
delayed, even more in the case of the full calculation, with hadronic
cascade. The bump in the two latter scenarios is due to particles
being produced from the fluid, the small $p_{t}$ contribution is
due to corona particles. 

This particular space-time behavior of the hydrodynamic expansions
should clearly affect Bose-Einstein correlations -- what we are going
to investigate in the following. %
\begin{figure}[b]
\begin{centering}
\includegraphics[angle=270,scale=0.32]{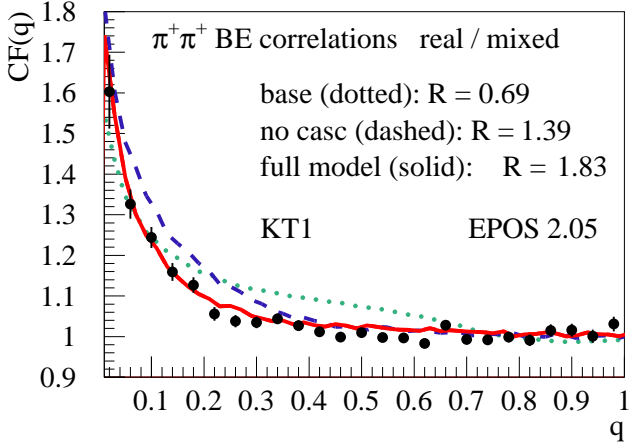}\caption{(Color online) The correlation functions $CF$ for $\pi^{+}$-- $\pi^{+}$
pairs as obtained from our simulations, for the three different scenarios,
for $k_{T}$ bin KT1, compared to data (points). \label{cap:cf}}

\par\end{centering}
\end{figure}
\begin{figure}[tb]
\begin{centering}
\includegraphics[angle=270,scale=0.32]{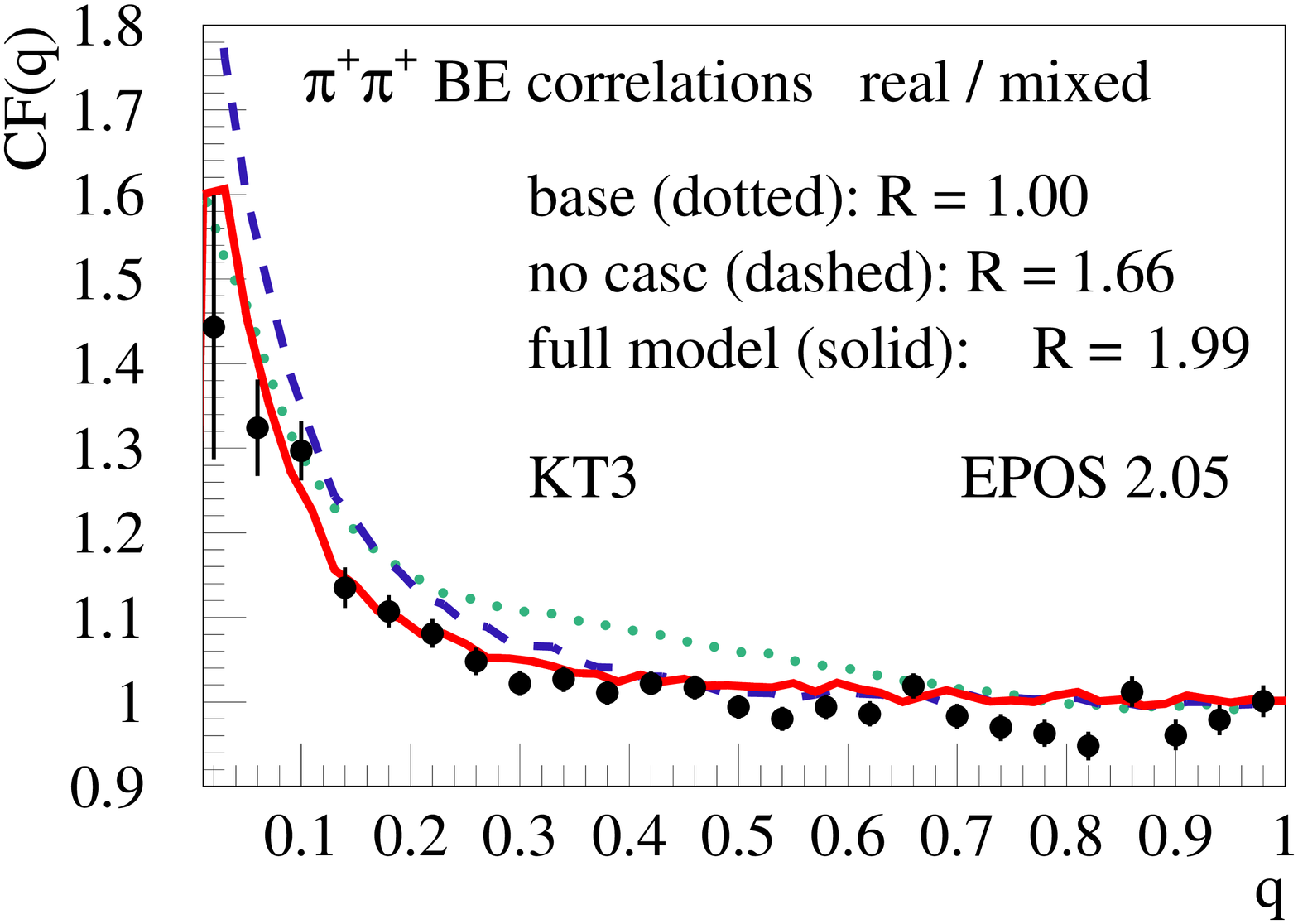}\caption{(Color online) Same as fig. \ref{cap:cf}, but $k_{T}$ range KT3.\label{cap:cf2}}

\par\end{centering}
\end{figure}
\begin{figure}[b]
\begin{centering}
\includegraphics[angle=270,scale=0.32]{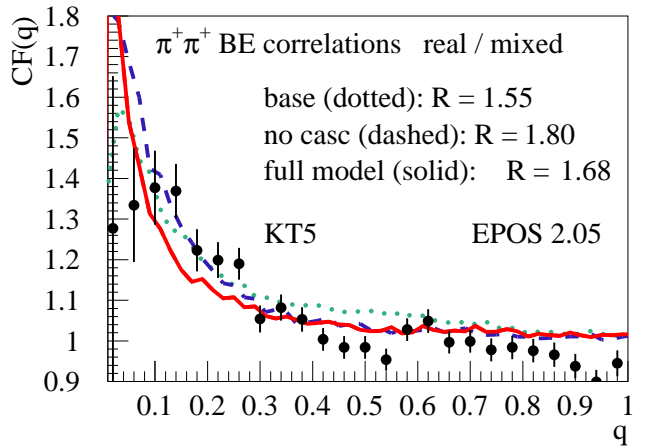}\caption{(Color online) Same as fig. \ref{cap:cf}, but $k_{T}$ range KT5.\label{cap:cf3}}

\par\end{centering}
\end{figure}
There is a long history of so-called femtoscopic methods \citet{hbt3,hbt4,hbt5,hbt6,hbt7},
where the study of two-particle correlations provides information
about the source function $S(\mathbf{P},\mathrm{\mathbf{r}'})$, being
the probability of emitting a pair with total momentum $\mathbf{P}$
and relative distance $\mathrm{\mathbf{r}'}$. Under certain assumptions,
the source function is related to the measurable two-particle correlation
function $CF(\mathbf{P},\mathbf{q})$ as \begin{equation}
CF(\mathbf{P},\mathbf{q})=\int d^{3}r'\, S(\mathbf{P},\mathbf{r}')\left|\Psi(\mathbf{q}',\mathbf{r}')\right|^{2},\label{eq:cf}\end{equation}
with $\mathbf{q}$ being the relative momentum, and where $\Psi$
is the outgoing two-particle wave function, with $\mathbf{q}'$ and
$\mathbf{r}'$ being relative momentum and distance in the pair center-of-mass
system. The source function $S$ can be obtained from our simulations,
concerning the pair wave function, we follow \citet{hbt-lednicki},
some details are given in \citet{epos2}. 

Here, we investigate $\pi^{+}$-- $\pi^{+}$ correlations. We evaluate
eq. (\ref{eq:cf}), with Bose-Einstein (BE) quantum statistics included,
but no Coulomb corrections. Weak decays are not carried out. In figs.
\ref{cap:cf}, \ref{cap:cf2},\ref{cap:cf3}, we show the results
for different $k_{T}$ intervals defined as (in MeV): KT1$=[100,250]$,
KT3$=[400,550]$, KT5$=[700,1000]$, where $k_{T}$ of the pair is
defined as \begin{equation}
k_{T}=\frac{1}{2}\left(|\vec{p}_{t}(\mathrm{pion}\,1)+\vec{p}{}_{t}(\mathrm{pion}\,2)|\right).\end{equation}
We compare the three different scenarios: {}``full calculation''
(solid line), {}``calculation without hadronic cascade'' (dashed),
and {}``calculation without hydro and without cascade'' (dotted),
and data from ALICE \citet{alice}. The data are actually not Coulomb
corrected, because the effect is estimated to be small compared to
the statistical errors. We consider here the high multiplicity class,
with $dn/d\eta(0)=11.2$, close to the value of 12.9 from our simulated
high multiplicity events. We compare with the real data (not polluted
with simulations), normalized via mixed events, and we do the same
with our simulations. Despite the limited statistics, in particular
at large $k_{T}$, we see very clearly that the {}``full'' scenario,
including hydro evolution and hadronic cascade, seems to fit the data
much better then the two other ones. Usually people like to extract
radii from these distributions, so when we make a fit of the form
\begin{equation}
CF-1=\lambda\,\exp(-R|\mathbf{q}|),\end{equation}
in the $|\mathbf{q}|$ range from 0.05 to 0.70. We obtain the radii
given in the figure. So the radii are very different, varying from
0.69 fm (base approach) to 1.80 fm (full model), which is understandable
from fig. \ref{cap:formation}. We prefer an exponential fit rather
than a Gaussian, simply because the former one works, the latter one
does not. We do not want to give a precise meaning to $R$, it simply
characterizes the distribution.

\begin{figure}[tb]
\begin{centering}
\includegraphics[angle=270,scale=0.32]{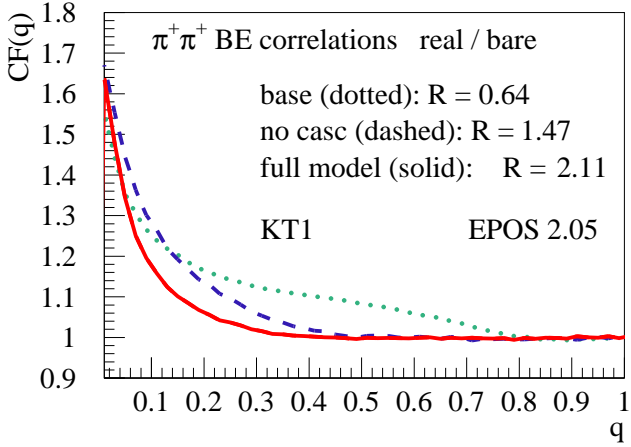}\caption{(Color online) Same as fig. \ref{cap:cf}, but normalization via a
simulation w/o BE correlation ({}``bare'').\label{cap:bare}}

\par\end{centering}
\end{figure}

\begin{figure}[b]
\begin{centering}
\includegraphics[angle=270,scale=0.32]{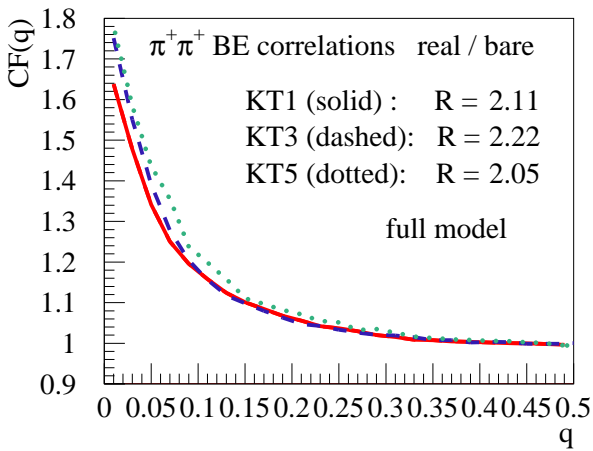}\caption{(Color online) Correlation function, normalized by using a simulation
w/o BE correlation ({}``bare''), for three $k_{T}$ intervals. \label{cap:full}}

\par\end{centering}
\end{figure}

Normalizing by mixed events is something one can easily do experimentally
(this is why we compare with these data), but it is clear that one
has still unwanted correlations, like those due to energy-momentum
conservation, which is not an issue in mixed events. Doing simulations,
life is easier. We can take simulations without Bose-Einstein correlations
as base line, rather than mixed events. This is referred to as {}``real
/ bare'' normalization (to be distinguished from the {}``real /
mixed'' case discussed earlier). The corresponding results are show
in fig. \ref{cap:bare}, the solid line (full calculation) is now
completely horizontal away from the peak region, the radius from the
exponential fit is 2.10 fm instead of 1.80 fm for the {}``mixed''
normalization. %
For the other $k_{T}$ regions, the situation is similar, the final
results for all three $k_{T}$ regions for the full calculation is
shown in fig. \ref{cap:full}, together with the the radii from the
exponential fit: they are almost identical, around 2 fm.%

We get to the same conclusion as outlined in \citet{alice}: the radii
are $k_{T}$ independent, contrary to what has been observed in AuAu
scattering.

\begin{figure}[tb]
\begin{centering}
\includegraphics[angle=270,scale=0.32]{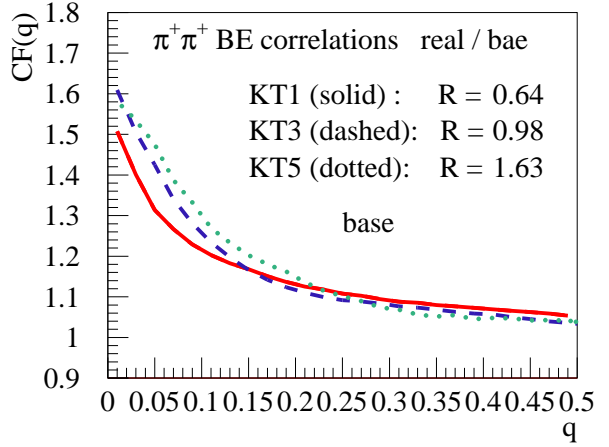}\caption{(Color online) As fig. \ref{cap:full}, but calculation without hydro
and without cascade. \label{cap:naive}}

\par\end{centering}
\end{figure}
\begin{figure}[b]
\begin{centering}
\includegraphics[angle=270,scale=0.32]{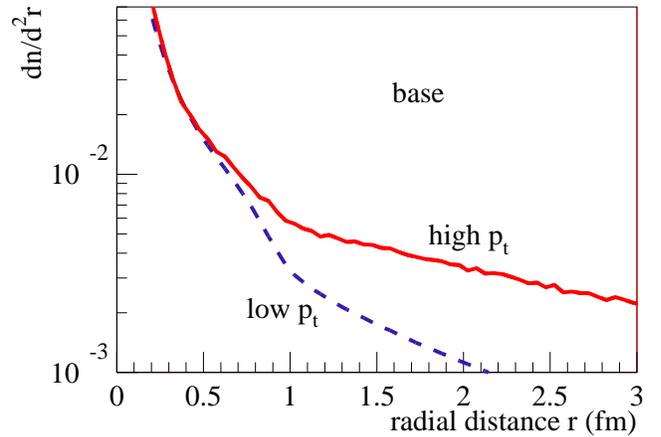}\caption{(Color online) The distribution of formation points of particles as
a function of the radial distance, for the scenario {}``without hydro
and without cascade'', for high $p_{t}$ and low $p_{t}$ particles.
\label{cap:low}}

\par\end{centering}
\end{figure}

How can it be that our hydrodynamic scenario gives a strong $k_{T}$
dependence in AuAu, but not in $pp$? To answer this question, we
compute the {}``true'' correlation function (real / bare normalization)
for the calculation without hydro and without cascade (just string
decay). The results are shown in fig. \ref{cap:naive}. Surprisingly,
here we get a strong $k_{T}$ dependence of the radii, but the {}``wrong''
way: we have 0.64 for KT1 and 1.63 fm for KT5 ! Actually such a behavior
is quite normal, as seen from fig. \ref{cap:low}: the distribution
is broader for high $p_{t}$ particles, because high $p_{t}$ resonances
live longer and can move further out before decaying. %
\begin{figure}[tb]
\begin{centering}
\includegraphics[angle=270,scale=0.32]{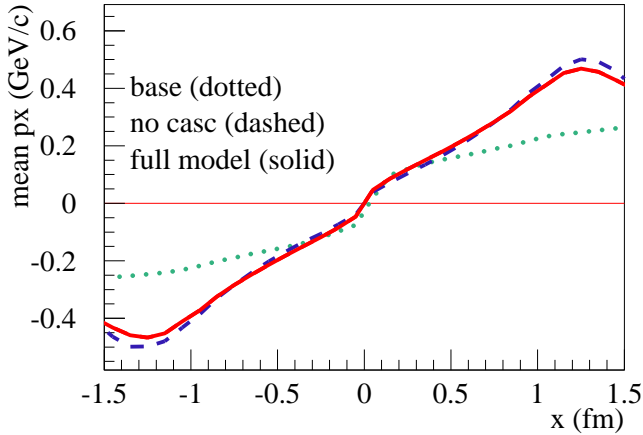}\caption{(Color online) Momentum -- space correlation, for the different scenarios.\label{cap:moment}}

\par\end{centering}
\end{figure}
This effect is in principle also present in AuAu scattering, but it
is much more visible for the small $pp$ system. So in $pp$ we have
two competing effects: 

\begin{itemize}
\item radii increase with $k_{T}$, due to the bigger size of the source
of the high $p_{t}$ particles compared to the low $p_{t}$ ones,
\item radii decrease with $k_{T}$, as in AuAu (see \citet{epos2}), in
case collective flow, due to the $p$-$x$ correlation. 
\end{itemize}
As seen in fig. \ref{cap:moment}, this $p-x$ correlation exists
indeed for the case of hydrodynamic evolutions, and is much smaller
in the basic scenario. %
\begin{figure}[b]
\begin{centering}
\includegraphics[angle=270,scale=0.32]{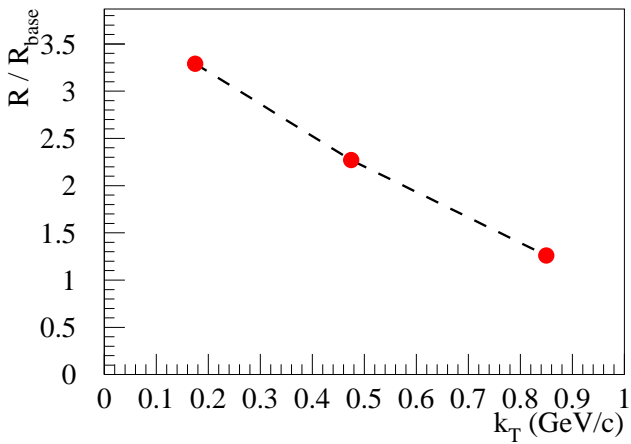}\caption{(Color online) The $k_{T}$ dependence of $R/R_{\mathrm{bas}}$. The
ratio decreases significantly with $k_{T}$, a clear {}``flow signal''.
\label{cap:kt}}

\par\end{centering}
\end{figure}
So in the hydro scenarios, the two competing effects roughly cancel,
the radii are $k_{T}$ independent. To really see the $x-p$ correlation,
one need to {}``divide out'' the trivial $k_{T}$ dependence due
to the $p_{t}$ dependence of the single particle source sizes, which
we do by considering the $k_{T}$ dependence of $R/R_{\mathrm{bas}}$,
with the reference radius $R_{\mathrm{bas}}$ referring to the base
scenario (without hydro, without cascade), see fig. \ref{cap:kt}:
The ratio $R/R_{\mathrm{bas}}$ decreases with $k_{T}$ as a manifestation
of the $x-p$ correlation, as a consequence of the hydrodynamic expansion. 

An alternative way of getting out unwanted correlations would be the
consideration of double ratios like\begin{equation}
\frac{CF(\mathrm{full\, scenario\, with\, BE})/CF(\mathrm{full\, scenario\, w/o\, BE})}{CF(\mathrm{base\, scenario\, with\, BE})/CF(\mathrm{base\, scenario\, w/o\, BE})},\end{equation}
where basic scenario refers to the calculation without hydro and without
hadronic cascade.

\section{Summary}

After having introduced recently a sophisticated approach of hydrodynamic
expansion based on flux-tube initial conditions for AuAu collisions
at RHIC, we now employ exactly the same picture to $pp$ scattering
at 900 GeV, which is in particular justified for high multiplicity
events. A very interesting application are Bose-Einstein correlations.
We have shown that as in heavy ion scattering the hydrodynamic expansion
leads to momentum -- space correlations, which clearly affect the
correlation functions. To see the signal is non-trivial due to the
fact that in addition to the $x-p$ correlations (which leads to decreasing
radii with $k_{T}$, there is a second effect which works the other
way round: the single particle source size is $p_{t}$ dependent,
which is an important effect in $pp$, not so in heavy ion scattering.
In this sense we can interpret the $k_{T}$ independence of the radii
as a real flow effect. Our simulation does not only reproduce the
$k_{T}$ independence, but also the whole correlation functions, which
is not at all reproduced from the {}``base scenario'' without hydro
and without cascade. So the correlation data provide a very strong
evidence for a collective hydrodynamic expansion in $pp$ scattering
at the LHC. 

\begin{acknowledgments}
This research has been carried out within the scope of the ERG (GDRE)
{}``Heavy ions at ultra-relativistic energies'', a European Research
Group comprising IN2P3/CNRS, Ecole des Mines de Nantes, Universite
de Nantes, Warsaw University of Technology, JINR Dubna, ITEP Moscow,
and Bogolyubov Institute for Theoretical Physics NAS of Ukraine. Iu.
K. acknowledges partial support by the MESU of Ukraine, and Fundamental
Research State Fund of Ukraine, agreement No F33/461-2009. Iu.K. and
K.W. acknowledge partial support by the Ukrainian-French grant {}``DNIPRO\char`\"{},
an agreement with MESU of Ukraine No M/4-2009. T.P. and K.W. acknowledge
partial support by a PICS (CNRS) with KIT (Karlsruhe). K.M. acknowledges
partial support by the RFBR-CNRS grants No 08-02-92496-NTsNIL\_a and
No 10-02-93111-NTsNIL\_a.
\end{acknowledgments}

\end{document}